\documentclass[aps,prl,floatfix,superscriptaddress,twocolumn,footinbib]{revtex4}
\usepackage{amssymb}
\usepackage{amsmath}
\usepackage{amsfonts}
\usepackage{appendix}
\usepackage{bm}
\usepackage{graphicx}
\usepackage{epsfig}
\usepackage{epstopdf}
\usepackage{balance}
\usepackage[dvipsnames]{xcolor}
\usepackage{calc}
\usepackage{natbib}
\usepackage[colorlinks,
            linkcolor=blue,
            anchorcolor=blue,
            citecolor=blue,
            urlcolor=blue]{hyperref}
\usepackage{lipsum}

\begin{document}
\title{Hydrogen-induced magnetism in graphene: a simple effective model description}
\author{Shuai Li}
\affiliation{School of Physics and Wuhan National High Magnetic Field Center,
Huazhong University of Science and Technology, Wuhan 430074,  China}
\author{Rui Yu}
\affiliation{School of Physics and Technology, Wuhan University, Wuhan 430072, China}
\author{Jin-Hua Gao}
\email{jinhua@hust.edu.cn}
\affiliation{School of Physics and Wuhan National High Magnetic Field Center,
Huazhong University of Science and Technology, Wuhan 430074,  China}
\author{X. C. Xie}
\affiliation{International Center for Quantum Materials, School of Physics, Peking University, Beijing 100871, China }
\affiliation{Collaborative Innovation Center of Quantum Matter, Beijing 100871, China}
\affiliation{CAS Center for Excellence in Topological Quantum Computation, University of Chinese Academy of Sciences, Beijing 100190, China}
\begin{abstract}Hydrogen adatoms induced  magnetic moment on graphene has been observed in atomic scale in a recent experiment [Gonz\'{a}lez-Herrero \textit{et} \textit{al}., Science 352, 437 (2016)]. Here, we demonstrate that all the experimental phenomena can be simply and well described by an equivalent Anderson impurity model, where  the electronic correlations on both  carbon and hydrogen atoms are represented by an effective on-site Coulomb interaction on hydrogen adatom $U_H$. This simple equivalent model works so well is because that the main effect of Coulomb interaction on both carbon and hydrogen atoms is just to split the impurity resonance states with different spin.  This effective Anderson impurity model can also well reproduce all the behaviours of hydrogen dimer on graphene observed in experiment. We thus argue that it should be a general model to describe the hydrogen adatoms on graphene with various adsorption configurations, especially for their magnetic properties. 
\end{abstract}
\maketitle
The adatom induced local magnetic moment in graphene is a very important topic, which has been intensively studied in last decade~\cite{Yazyev2007,kats2008,Ulybyshev2015,Duplock2004,Anomhydrograph,science2016,nanoletter09,balseiro2012,Sahin2009,Wehling2009,EmbTranAtom,Yazyev2010,zuquanzhang2017,FerroNitrograph,Kim2013,Nokelainen2019,Charlebois2015,Haase2011,Yazyev2008,Ugeda2010,Hadipour2019,Pike2014,Santos2012}. It is of special importance for the application of graphene in spintronics~\cite{Han2014Graphene,Rocha2010,Soriano2010}. A typical example is the hydrogen adatom on graphene~\cite{Yazyev2007,Duplock2004,Anomhydrograph,science2016,nanoletter09,kats2008,balseiro2012,Ulybyshev2015,Sahin2009}. It is predicted by the first principles calculations that a hydrogen atom favors the top site of graphene, and can generate a local magnetic moment around the adsorption site~\cite{Duplock2004,Yazyev2007}. More excitingly, in a recent experiment, a single hydrogen adatom induced magnetic moment has been observed by scanning tunneling microscopy (STM)~\cite{science2016}.

In the experiment of Ref.~\onlinecite{science2016}, the most important experimental observation is a hydrogen adatom induced spin-split resonance state on graphene. Around the hydrogen adatom, the STM measurements find two narrow resonance peaks near the Fermi level, which are separated in energy by a splitting about $20$ meV. The first principle calculations show that the two resonance peaks are spin polarized and correspond to a spin-split resonance state, which gives rise to a local magnetic moment on graphene. One unique characteristic of this spin-split resonance state is that it  only locates on the graphene sublattice which is not directly coupled to the hydrogen atom.
Commonly, it is believed that a hydrogen adatom is equivalent to a carbon vacancy, which will induce a resonance state on graphene. And, due to the Coulomb interaction on carbon atoms ($U_C$), this resonance state becomes spin-splitting and thus induces a local magnetic moment on graphene~\cite{Yazyev2007,Yazyev2010}.

So far, to theoretically describe such hydrogen adatom induced magnetism on graphene, it needs to consider the Coulomb interaction on carbon atoms. It makes the calculations rather complicated.  The two possible ways are the DFT simulation and the Hubbard model. The two methods can only give some numerical simulations. We actually still do not have  a simple model which can give some analytical understanding about the hydrogen adatom induced magnetism. 

In this work, we propose a simple equivalent Anderson impurity model to describe the hydrogen adatom induced  magnetism  on graphene. 
All the experimental phenomena in Ref.~\onlinecite{science2016} can be simply and well described by this effective model, where the influence of the Coulomb interaction on both carbon and hydrogen atoms are attributed to a single  parameter $U_H$, i.e. the effective (or equivalent) Coulomb interaction on the hydrogen adatom. This effective model works so well is because the fact that the main role of the Coulomb interaction on graphene  ($U_C$) here is just to break the spin degeneracy of the impurity resonance state.  Note that,  $U_H$ is  able to break the spin degeneracy of the  resonance state on graphene as well. Thus, equivalently, we can use an effective $U_H$ to phenomenologically mimic the electronic correlations on both carbon and hydrogen atoms.  The advantage of this effective model is its simplicity, with which the computation complexity has been greatly simplified and we can even get some analytical understanding about the induced magnetism on graphene. 
We also calculate the magnetic properties of hydrogen dimer on graphene with this effective model. The results  coincide well with the experimental observations. Thus, we argue that this effective Anderson impurity model offers a simple way to understand the behaviors of the hydrogen adatoms on graphene with various adsorption configurations, especially for their magnetic properties.

We first show that an Anderson impurity can give rise to a spin-spilt resonance state on graphene. Here, the hydrogen adatom is described as an impurity.  The total Hamiltonian is $H=H_0+H_{\textrm{imp}}+H_{\textrm{hop}}$.
$H_0$ is the Hamiltonian of graphene
\begin{equation}
H_0=-t\sum_{\mathbf{k},\sigma}[\phi(\mathbf{k})a^{\dagger}_{\mathbf{k}\sigma}b_{\mathbf{k}\sigma}+\phi^*(\mathbf{k})b^{\dagger}_{\mathbf{k}\sigma}a_{\mathbf{k}\sigma}]
\end{equation}
where $a_{\mathbf{k}\sigma}$ ($b_{\mathbf{k}}$) is the annihilation operator of the graphene electrons on A (B) sublattice.  $\mathbf{k}$ and $\sigma$ are the moment and spin index, respectively. $t$  is the nearest neighbor hopping. $\phi(\mathbf{k})=\sum_je^{i \mathbf{k} \cdot \mathbf{\delta}_j}$, where $\delta_j$ is the nearest neighbour vectors. The adatom is described as an Anderson impurity,
\begin{equation}
H_{\textrm{imp}}=\sum_{\sigma}\epsilon_0d^{\dagger}_\sigma d_{\sigma}+U_H d^{\dagger}_\uparrow d_{\uparrow}d^{\dagger}_\downarrow d_{\downarrow}.
\end{equation}
Here, $\epsilon_0$ is the energy level of the impurity state, $d^\dagger_\sigma$ is the corresponding creation operator of the impurity electron with spin $\sigma$, and $U_H$ is the on-site Coulomb interaction. We assume that the adatom is on top of one carbon atom  in sublattice A, which is just the case of hydrogen adatom. The hopping term is
\begin{equation}
H_{\textrm{hop}}=\frac{V}{\sqrt{N}}\sum_{\mathbf{k}\sigma}[d^\dagger_\sigma a_{\textbf{k}\sigma} + \textrm{H}.\textrm{c}.],
\end{equation}
where $N$ is the number of lattice sites in one sublattice.

\begin{figure}[t!]
\centering
\includegraphics[width=8.5cm]{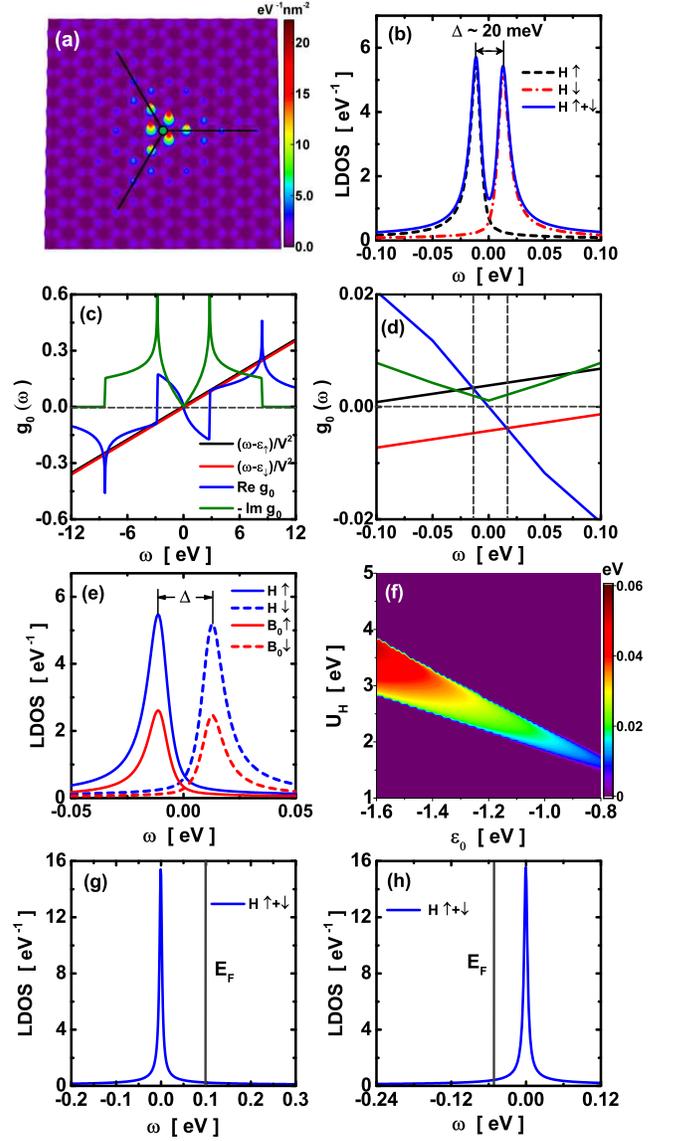}
\caption{(a) The  LDOS pattern of graphene with an adsorbed H adatom  at energy $\omega=0.2$ eV. (b) LDOS at the H adatom. (c) and (d) The graphical solutions of Eq.~\eqref{impurityposition} in different energy region. (e) LDOS at carbon atom $\textrm{B}_0$, where position of $\textrm{B}_0$ is given in Fig.~\ref{fig2}. The LDOS at H are also plotted as comparison.  (f) The energy separation $\Delta$  as a function of $\varepsilon_0$ and $U_H$£¬¡¡ with $V=5.8$ eV and $E_F=0$ eV.  (g)  LDOS at H adatom when graphene is electron doped ($E_F=0.1 $eV) and (h) hole doped ($E_F=-0.05 $eV). The parameters: $\epsilon_0=-1.18$ eV, $U_H=2.4$ eV, $V=5.8$ eV.}
\label{fig1}
\end{figure}

\begin{figure*}[t!]
\centering
\includegraphics[width=18cm]{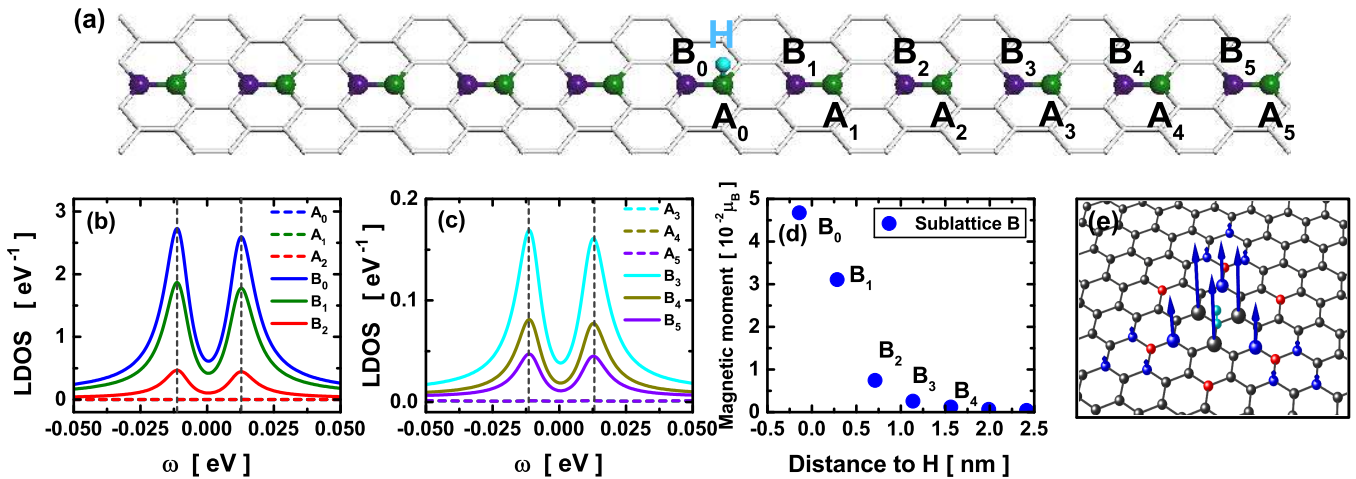}
\caption{(a) Schematic of the graphene structure along the solid line in Fig.~\hyperref[fig1]{1(a)}.
  (b) and (c) are the LDOS at  carbon atoms. (d) and (e) are the local magnetic moments at carbon atoms. Blue (red) arrows are for spin up (down)¡¡magnetic moment. The parameters: $\varepsilon_0=-1.18$ eV, $U_H=2.4$ eV, $V=5.8$ eV.}
\label{fig2}
\end{figure*}

The on-site Coulomb interaction is treated with mean field approximation
\begin{equation}
U_Hd^{\dagger}_\uparrow d_{\uparrow}d^{\dagger}_\downarrow d_{\downarrow} \approx U_H [d^{\dagger}_\uparrow d_{\uparrow}\langle n_{\downarrow}\rangle + d^{\dagger}_\downarrow d_{\downarrow} \langle n_{\uparrow}\rangle  - \langle n_{\downarrow}\rangle \langle n_{\uparrow}\rangle ].
\end{equation}
We define $\epsilon_\sigma=\epsilon_0 + U_H \langle n_{\bar{\sigma}} \rangle$. And $\langle n_\sigma \rangle = \langle d^{\dagger}_\sigma d_{\sigma} \rangle$  is the occupation of the impurity state with spin $\sigma$, which can be solved self-consistently
\begin{equation}\label{eqnd}
\langle n_\sigma \rangle =  \int^{\mu} _ {-\infty} d\omega \rho_{\textrm{H}\sigma}(\omega).
\end{equation}
Here, $\rho_{\textrm{H}\sigma}(\omega)=-\frac{1}{\pi}\textrm{Im}G_{dd,\sigma}(\omega)$ is the LDOS of the adatom.  $G_{dd,\sigma}(\omega)$ is  the retarded Green's function of the adatom
\begin{equation}\label{impuritygf}
G_{dd,\sigma}(\omega)=\frac{1}{\omega-\epsilon_\sigma-V^2 g_{00}(\omega)+i 0^+},
\end{equation}
where $g(\omega)=(\omega-H_0+ i0^+)^{-1}$ is the retarded Green's function of pristine graphene.  $g_{ii}(\omega)$ is the diagonal matrix element of $g(\omega)$, which describes the $i$-th carbon atom. For example, $g_{00}(\omega)$ is for the host carbon atom which is coupled to the hydrogen adatom, and actually $g_{ii}(\omega)=g_{00}(\omega)$ for pristine graphene. To calculate the retarded Green's function of the graphene,  the Dyson equation is
\begin{equation}\label{graphenegf}
G_{ii,\sigma}(\omega)=g_{ii}(\omega) + g_{i0}(\omega)T_{00,\sigma}(\omega)g_{0i}(\omega)
\end{equation}
where $T_{00,\sigma}(\omega)=\frac{V^2}{\omega-\epsilon_\sigma-V^2g_{00}(\omega)+i0^+}$  is an element of t-matrix.
 So, the LDOS of the $i$-th site of graphene is
\begin{equation}\label{grapheneldos}
\rho_{i,\sigma}(\omega)=-\frac{1}{\pi}\textrm{Im}G_{ii,\sigma}(\omega).
\end{equation}
Note that, once $\langle n_{\sigma} \rangle$ is got by solving Eq.~\eqref{eqnd}, $\epsilon_\sigma$ is known and then we can directly calculate the LDOS of carbon atom£ó in graphene with Eqs.~\eqref{graphenegf} and \eqref{grapheneldos} [See in Fig.~\hyperref[fig1]{1(a)}].

The magnetic phase diagram of the adatom on graphene has been intensively studied~~\cite{zuquanzhang2017,uchoa2008,Uchoa2014,McCreary2012,Mashkoori2015}.  With proper parameters, the adatom will become magnetic with $\langle n_{\uparrow} \rangle \neq \langle n_{\downarrow} \rangle$. In this case, due to  $U_H$,  we get two spin split impurity levels on hydrogen adatom, where one is below  and the other is above the Fermi level [See in Fig.~\hyperref[fig1]{1(b)}]. Numerically, the positions of the two impurity levels are determined  by setting the denominator in Eq.~\eqref{impuritygf} equal to zero
\begin{equation}\label{impurityposition}
\omega-\epsilon_\sigma-V^2 \textrm{Re} [g_{00}(\omega)]=0,
\end{equation}
while $V^2 \textrm{Im}[g_{00}(\omega)]$ gives the width of the impurity levels.
The graphical solutions of Eq.~\eqref{impurityposition} for each spin are shown in Figs.~\hyperref[fig1]{1(c)} and \hyperref[fig1]{1(d)}, and LDOS of hydrogen atom $\rho_{\textrm{H}\sigma} (\omega)$ is plotted in Fig.~\hyperref[fig1]{1(b)}. The two spin split impurity peaks are shown clearly.

Now, we discuss the  resonance states on graphene. Once the adatom becomes magnetic, the adatom is equivalent to a spin-dependent potential $V^2/(\epsilon-\epsilon_\sigma)$ applied on the host carbon atom. So, the induced resonance states with the up spin and down spin are no longer degenerate. Numerically, the resonance states correspond to the poles of the t-matrix given in Eq.~\eqref{graphenegf}, the positions of which are thus also determined by Eq.~\eqref{impurityposition}. It means that the energy positions of resonance states on graphene are the same as that of the impurity levels on hydrogen adatom. 
Note that, to guarantee the adatom to be magnetic, we always have one impurity level ($\textit{e}.\textit{g}.$ spin down) above  and the other ($\textit{e}.\textit{g}.$ spin up) below $E_F$.  Correspondingly, as shown in Fig.~\hyperref[fig1]{1(e)}, the resonance state with spin up is always below, and  the one with spin down is always above $E_F$.
 Therefore, we see that an Anderson impurity will intrinsically  induce a spin polarized resonance state on graphene, which is the result of $U_H$. The unpaired electron in the spin-split resonance state gives rise to a local magnetic moment. 
 
 It should be noted that the two spin polarized resonance peaks (and the impurity levels) are asymmetric. For the impurity levels, it is obvious because that  their peak broadening $V^2 \textrm{Im}[g_{00}(\omega)]$  are different due to the energy difference [See Fig.~\hyperref[fig1]{1(d)}].  As shown in Figs.~\hyperref[fig1]{1(b)} and \hyperref[fig1]{1(e)}, our calculations show that the impurity peak above $E_F$ is  lower than the one below $E_F$. The two resonance peaks have similar behaviour. 
  The next nearest neighbor hopping in graphene $t'$ may further
  enhance the asymmetric of the two resonance peaks, because that $t'$ will break the particle-hole symmetry of graphene, \textit{i.e.}, $\textrm{Im}[g_{00}(\omega)]$ becomes asymmetry  now.



In reality, $U_H$ and $U_C$ always coexist.
We can  turn on $U_H$ and $U_C$ in sequence to understand the influence of $U_C$. 
Considering $U_H$ first, 
the adatom is an Anderson impurity, which will be magnetic or nonmagnetic depending on the the interplay of  $U_H$, $V$, $\epsilon_0$ and doping~\cite{uchoa2008}.
Then, we turn on $U_C$. Intuitively, the only effect of $U_c$ we can expect is to further enhance the spin splitting of the resonance state on graphene, or induce the spin splitting if $U_H$ is not large enough to break the spin degeneracy of the impurity level. Correspondingly,  the impurity levels of the hydrogen adatom should be modified by $U_C$ as well.
 Note that $U_C$ can not significantly  change the magnitude of the induced  magnetic moment. It is because that, no matter what value $U_C$ is, we always and only have a spin polarized resonance state below $E_F$. 

Based on the above discussions, it is not difficult to expect that we can approximately  use a renormalized $U_H$  to represent the electronic correlation on both hydrogen and carbon atoms.  Here, $U_H$ should be renormalized in order to account for the enhanced spin splitting of the resonance states due to $U_C$, while $\epsilon_0$ is modefied by $U_C$  as well. It means that, in fact, we can use an effective Anderson impurity model to describe the hydrogen adatom induced spin-split resonance states on graphene, which is central result of our work.
 In the following, we first get the  values of effective $U_H$ and $\epsilon_0$ by fitting the experiment, and then we will show that all the experimental observations in Ref.~\cite{science2016} can be well described by this effective model.


In experiment, the STM measurement indicates that the energy separation of the two impurity levels $\Delta$ [See in Fig.~\hyperref[fig1]{1(b)}] is about 20 meV~\cite{science2016}.  In the Anderson impurity model, there are three key parameters: the energy of the impurity level $\epsilon_0$, the hybridization between the impurity and the carbon atom $V$ and the on-site Coulomb interaction of the adatom $U_H$. Compared with the DFT calculations, it is generally accepted that $V$ is about $5.8$ eV. In this effective model, $\epsilon_0$  is  not its bare value, but an effective parameter to include the influence of $U_C$ and other realistic details, $\textit{e}.\textit{g}.$ the lattice distortion around the adatom. $U_H$ is also an effective parameter to be determined. 
  In Fig.~\hyperref[fig1]{1(f)}, we plot $\Delta$ as a function of $\epsilon_0$ and $U_H$. Considering the experimental result $\Delta \approx 20$ meV, we get the effective parameters: $\epsilon_0 \approx -1.18$ eV, $U_H \approx 2.4$ eV.
The choice of the value of effective parameters is not unique, but in a small parameter region as shown in Fig.~\hyperref[fig1]{1(f)}. The two effective parameters above can give a rather good description about the hydrogen adatoms on graphene.

\begin{figure}[t!]
\centering
\includegraphics[width=8.5cm]{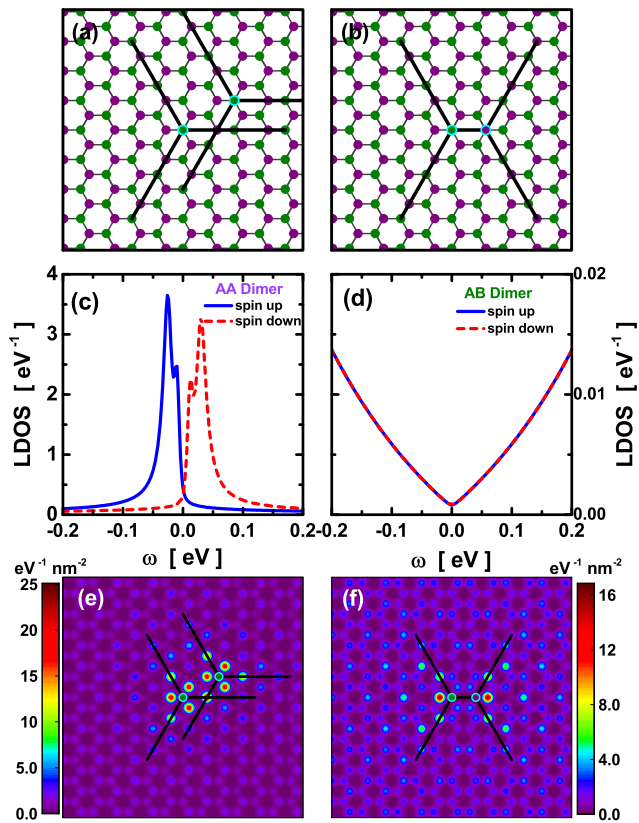}
\caption{(a) and (b) are AA and AB dimer configurations, respectively. (c) and (d) are the LDOS at H adatoms for AA and AB dimer, respectively.   (e) and (f) are the LDOS pattern at $\omega=0.2$ eV for  AA and AB dimer, respectively.  Parameters are: $\varepsilon_0=-1.18$ eV, $U=2.4$ eV, $V=5.8$ eV, and $E_f=0$ eV.}
\label{fig3}
\end{figure}

This effective model can well interpret the doping effect on the local magnetic moment. With a large enough charge (hole) doping, the adatom can be changed from the magnetic phase into nonmagnetic phase,  which causes the splitting of the hydrogen energy levels (and that of the resonance states) to vanish [See in Figs.~\hyperref[fig1]{1(g)} and \hyperref[fig1]{1(h)}].

This effective model can well describe the spin spilt resonance states and the  local magnetic moment induced by the hydrogen adatom.
 As shown in Fig.~\hyperref[fig1]{1(a)}, when a hydrogen atom is adsorbed, the LDOS of graphene exhibits a 3-fold symmetry with three ``arms" at $120^{\circ}$ each (see the three black solid lines). To compare with the experiment, we plot the LDOS of carbon atoms along one of the arms [See Fig.~\hyperref[fig2]{2(a)}] in Figs.~\hyperref[fig2]{2(b)} and \hyperref[fig2]{2(c)}. Here, the hydrogen adatom is on A sublattice.  We see that the LDOS at all the carbon atoms on A sublattice is nearly zero (dashed lines), while obvious LDOS peaks corresponding to the two resonance states (solid lines) have been observed on all the carbon sites of the B sublattice. The sublattice asymmetry in the spatial distribution is a common feature for such kind of resonance state induced by local perturbation on one carbon atom~\cite{Lawlor2013}.
 As indicated in Fig.~\hyperref[fig1]{1(e)}, the resonance peaks on carbon atoms are spin dependent, which means that there is a spatial distribution of local magnetic moment on B sublattice and nearly zero local magnetic moment on sublattice A. We  plot the corresponding local magnetic moment on carbon atoms in Figs.~\hyperref[fig2]{2(d)} and \hyperref[fig2]{2(e)}. With the effective parameters we choose, the calculated local magnetic moments on carbon atoms [see Fig.~\hyperref[fig2]{2(d)}] are  close to the DFT results ~\cite{science2016}. Interestingly, this simple effective model suggests that the local magnetic moment around the hydrogen adatom [see Ref.~\onlinecite{science2016} and Fig.~\hyperref[fig2]{2(d)}] can be viewed as a special spin-dependent Friedel oscillation at half filling with a $r^{-3}$ decay. The detailed analytical proof will be given elsewhere~\cite{notice}.

This effective model can well interpret the magnetic property of hydrogen dimer.   We view the hydrogen dimer as  two Anderson impurities on graphene, and use the same effective parameters. Here,  the configurations of AA and AB dimers are the same as that in the experiment, as illustrated in Figs.~\hyperref[fig3]{3(a)} and \hyperref[fig3]{3(b)}.
By the self-consistent mean field method, we can determine that whether the two adatoms are magnetic or not. The results are plotted in Fig.~\ref{fig3}. For a AA dimer, we calculate the LDOS at the hydrogen atom [See in Fig.~\hyperref[fig3]{3(c)}]. For the up spin, there is a resonance peak with fine structure (an additional tiny peak appears on the shoulder of the resonance peak) below $E_F$, while a similar resonance peak with down spin is found above the $E_F$. Thus, our model predicts that the hydrogen AA dimer is magnetic, and the calculated energy splitting is about 50 meV.
The LDOS of AB dimer is given in Fig.~\hyperref[fig3]{3(d)}. The hydrogen adatoms are now  in nonmagnetic phase, and there is no resonance peaks near the Fermi level.  The LDOS pattern of the AA and AB dimer are shown in Figs.~\hyperref[fig3]{3(e)} and \hyperref[fig3]{3(f)}, respectively.  These results coincide  well with the experimental observation as well as DFT simulations~\cite{kats2008,science2016}.


In short, we point out that the recent experiment about the hydrogen adatoms on graphene can be well described by an effective Anderson impurity model.  It can also give an good description about the hydrogen dimmer. We thus argue that it is a general theoretical model to understand the behaviors of hydrogen adatoms on graphene with various adsorption configurations. The hydrogen adatoms on graphene may be an ideal platform to study the Anderson impurities in atomic scale.

\begin{acknowledgments}
This work is funded by the National Key Research and Development Program of China
(Grants No. 2017YFA0403501 and No. 2016YFA0401003), and National Natural Science Foundation of China (Grants No. 11534001, 11874160, 11274129).
\end{acknowledgments}

\bibliographystyle{apsrev4-1}
\bibliography{hatombib}

\end{document}